\documentclass[12pt]{article}

\usepackage[utf8]{inputenc}
\usepackage[english]{babel}
\usepackage{amsmath,amsfonts,amssymb,amsthm}
\usepackage{lmodern}
\usepackage[scaled]{helvet}
\usepackage[T1]{fontenc}
\usepackage{lettrine} 
\usepackage{graphicx,epstopdf,color}
\usepackage{booktabs}
\usepackage{authblk}

\RequirePackage{fullpage}
\RequirePackage[colorlinks=true,citecolor=blue,menucolor=black,linkcolor=black]{hyperref}
\usepackage[section]{placeins}
\usepackage{subcaption}
\usepackage{float}

\usepackage{parskip}

\begingroup
    \makeatletter
    \@for\theoremstyle:=definition,remark,plain\do{%
        \expandafter\g@addto@macro\csname th@\theoremstyle\endcsname{%
            \addtolength\thm@preskip\parskip
            }%
        }
\endgroup

\usepackage{color}
\usepackage{ifthen}

  \newcommand{\R}{\ensuremath{\mathbb{R}}}
  \newcommand{\E}{\ensuremath{\mathbb{E}}}
  
  \newcommand{\Nc}{\mathcal{N}}

  \newcommand{\Mc}{\mathcal{M}}

  \newcommand{\Ac}{\mathcal{A}}

\newcommand{\norm}[1]{\left|\left| #1 \right|\right|}

\newcommand{\abs}[1]{\left \vert #1 \right \vert}

\newcommand{\TODO}[1]{ 
  \ifx\NOTES\undefined\else
	{\tt \color{red} [TODO:#1] } 
  \fi
}

\newcommand{\TODOM}[1]{ 
  \ifx\NOTES\undefined\else
  {\tt \color{green} [TODO FOR MATAN:#1] } 
  \fi
}

\newcommand{\NOTE}[1]{ 

\ifx\NOTES\undefined\else
  \footnote{ {\color{blue} NOTE: #1}}  
\fi
}

\newcommand{\ecomment}[1]{ 
\ifx\NOTES\undefined\else 
{\color{blue}[E]}\footnote{ {\color{blue} Eran: #1}}
\fi
}

\newcommand{\mcomment}[1]{ 
\ifx\NOTES\undefined\else
  {\color{green} [M]}\footnote{ {\color{green} Matan: #1}}  
\fi
}

\newcommand{\Pd}[3]{\ifthenelse{\equal{#3}{1}}{\frac{\partial #1}{\partial #2}}{\frac{\partial^{#3} #1}{\partial #2^{#3}}}}

\newcommand{\T}{\ensuremath{\top}}
\newcommand{\AMPSVST}{{\bf AMP-SVST}}
\newcommand{\AMPOPT}{{\bf AMP-OPT}}
\newcommand{\TW}{{\bf NIHT}}
\newcommand{\TY}{{\bf APG}}
\newcommand{\nnm}{{nnm}}
\newcommand{\IT}{{it}}

\newcommand{\qref}[1]{({\ref{#1}})}

\title{Near-optimal matrix recovery from random linear measurements}
\author[1]{Elad Romanov}
\author[1]{Matan Gavish \thanks{To whom correspondence should be addressed. E-mail: gavish@cs.huji.ac.il}} 
\affil[1]{School of Computer Science and Engineering, The Hebrew University, 
Jerusalem, Israel}

\date{}

\begin{document}

\maketitle

\begin{abstract}

In matrix recovery from random linear measurements, one is interested in
recovering an unknown $M$-by-$N$ matrix $X_0$ from $n<MN$ measurements
$y_i=Tr(A_i^\T X_0)$ where each $A_i$ is an $M$-by-$N$ measurement 
matrix with i.i.d random entries,
$i=1,\ldots,n$.
We present a novel matrix recovery algorithm, based on approximate message
passing, which iteratively  applies an optimal singular value shrinker -- a
nonconvex nonlinearity tailored specifically for matrix estimation.  Our
algorithm typically converges exponentially fast, offering a significant speedup
over previously suggested matrix recovery algorithms, such as iterative solvers
for Nuclear Norm Minimization (NNM).  It is well known that there is a recovery
tradeoff between the information content of the object $X_0$ to be recovered
(specifically, its matrix rank $r$) and the number of linear measurements $n$
from which recovery is to be attempted.  
The precise tradeoff between $r$ and $n$, beyond which recovery by a given
algorithm becomes possible, traces the so-called phase transition curve of that
algorithm in the $(r,n)$ plane.  The phase transition curve of
our algorithm is 
noticeably better than that of NNM. Interestingly, it is close to the
information-theoretic lower bound for the minimal number of measurements needed
for matrix recovery, making it not only state-of-the-art in terms of convergence
rate, but also
near-optimal in terms of the matrices it successfully recovers.

\end{abstract}

Modern datasets often take the form of large matrices.  When the full
dataset is not directly observable, the scientist can only obtain measurements,
from which she hopes to recover the dataset.  
Let $X$ be our $M$-by-$N$ data
matrix, and consider $n$ linear measurements of $X$,
$y_i=Tr(A_i^\T X)$ ($i=1,\ldots,n$), where  $A_1,\ldots,
A_n$ are measurement matrices.  Basic linear algebra tells us that, in order to reconstruct $X$ from
these measurements using a linear algorithm, successful recovery is only
possible if $n\geq MN$.  In recent years, intense research in applied
mathematics, optimization, and information theory has shown that, when the rank
$r=rank(X)$ is low, nonlinear algorithms based on convex optimization allow
exact recovery of $X$ from  only
$O(rM+rN)$ measurements, up
to logarithmic factors, 
whereby solving a severely underdetermined system of linear
equations.  In recent years, research in matrix recovery has flourished spanning 
both theory \cite{candes2012exact,candes2010power,gross2011recovering,keshavan2010matrix,recht2010guaranteed,candes2010matrix} and efficient algorithms \cite{cai2010singular,ma2011fixed,toh2010accelerated,jain2010guaranteed,tanner2013normalized,jain2013low}.  
Applications have been developed in fields ranging
widely, from video and image processing \cite{candes2011robust,zhou2015low} and system identification \cite{liu2009interior,recht2010guaranteed} to quantum state
tomography \cite{gross2010quantum} and collaborative
filtering \cite{keshavan2010matrix,candes2012exact};
see also
the recent survey \cite{davenport2016overview}.

It is convenient to denote the linear measurement scheme by
a linear operator $\Ac:\R^{M\times N} \to \R^n$, so that $y=\Ac(X)$ are
the $n$ linear measurements available to the scientist.
As a simple model for random linear measurements, one takes $A_i$ to be white
Gaussian vectors, or equivalently, assumes that the entries of $\Ac$ are
independent and identically distributed $\Nc(0,1/n)$ random variables.
These are known as Gaussian measurements; more generally we will assume that the
entries of $\Ac$ are i.i.d from some zero-mean distribution with variance
$1/n$, and refer to this model simply as random linear measurements.

As matrix recovery problems contain $MN$ degrees of freedom, 
their size grows quickly with $M$ and $N$,
making direct solvers infeasible; 
nearly all
existing algorithms approach the problem iteratively.
In fact, most matrix recovery algorithms
proceed by some variation of 
 iterative singular value shrinkage 
 \begin{eqnarray} \label{GD:eq}
	X_{t+1} &=&  \eta_t(X_t + \mu_t \Ac^* z_t)\\
	z_t &=&  y - \Ac X_t \,,\nonumber
 \end{eqnarray}
 where $X_t$ is the current estimate for $X$, $\mu_t$ is a step size and 
 $z_t\in\R^n$ is the current residual vector. Here,
 $\eta_t:\R^{M\times N}\to \R^{M\times N}$ is a {\em singular value shrinker},
 namely,
 a matrix function that applies the same univariate nonlinearity 
 to each of the
 singular values of its matrix argument. 
 Abusing notation, here and below we write $\eta_t$ both 
 for the univariate nonlinearity and for the corresponding matrix function.

 In this paper we propose two new matrix recovery algorithms, based on the
 Approximate Message Passing (AMP) framework \cite{donoho2009message}.
 These algorithms offer significant improvement over variations of 
 \qref{GD:eq}.
 Starting at $X_0=0$, we propose the Matrix AMP iteration  
 \begin{eqnarray} \label{AMP:eq}
	X_{t+1} &=&  \eta_t \left( X_t + \Ac^* z_t \right) \\
	z_t &=&  y - \Ac X_t + b_t z_{t-1} \nonumber
\end{eqnarray}
where $\eta_t$ are a sequence of singular value denoisers and 
\begin{equation}
	b_t = 
	\frac{1}{n} \nabla \cdot \eta_{t-1}
	\left( X_{t-1} + \Ac^* z_{t-1} \right) \,.
\end{equation}
Here, $\nabla \cdot \eta_t$ is the divergence of the matrix function $\eta_t$.
These formulas are a very natural extension 
of the AMP framework, originally
developed for recovery of sparse vectors, to the matrix recovery problem.
Any single singular value shrinker $\eta$ can yield a full-blown AMP matrix
recovery algorithm by setting 
\begin{eqnarray} \label{eta:eq}
       \eta_t(W) = \hat{\sigma}_t \eta(W/\hat{\sigma}_t)
\end{eqnarray}
in \qref{AMP:eq},
where $\hat{\sigma}_t$ is the current
noise  level estimate rigorously defined below. 
Comparison of Eqs. (\ref{GD:eq}) and (\ref{AMP:eq}) reveals a subtle 
yet crucial
difference
between the two algorithms, known as the {\em Onsager correction term}, 
which is extensively discussed in the AMP
literature
\cite{donoho2009message,donoho2011noise,donoho2013accurate,montanari2012graphical,BARA}. 

The first algorithm we present, \AMPSVST,
is a Matrix AMP iteration based on the famous 
soft thresholding nonlinearity
\begin{eqnarray} \label{soft:eq}
       \eta^{soft}_\lambda(y) = (y-\lambda)_+
\end{eqnarray}
with a tuning parameter that depends on $M$,$N$ and $r$ in a manner described
below. The second algorithm we present, \AMPOPT, is a Matrix AMP iteration based on 
a variation of the asymptotically 
optimal singular value shrinker for low-rank matrix denoising
\cite{gavish2014optimal}. The explicit form of
shrinker, which we denote by $\eta^{opt}$, is based on \qref{opt:eq} and 
\qref{eta_opt:eq} below. Figure
\ref{shrinkers:fig} compares the two shrinkers with the hard thresholding
nonlinearity.

\begin{figure}[h!]
       \centering
       \includegraphics[width=\columnwidth]{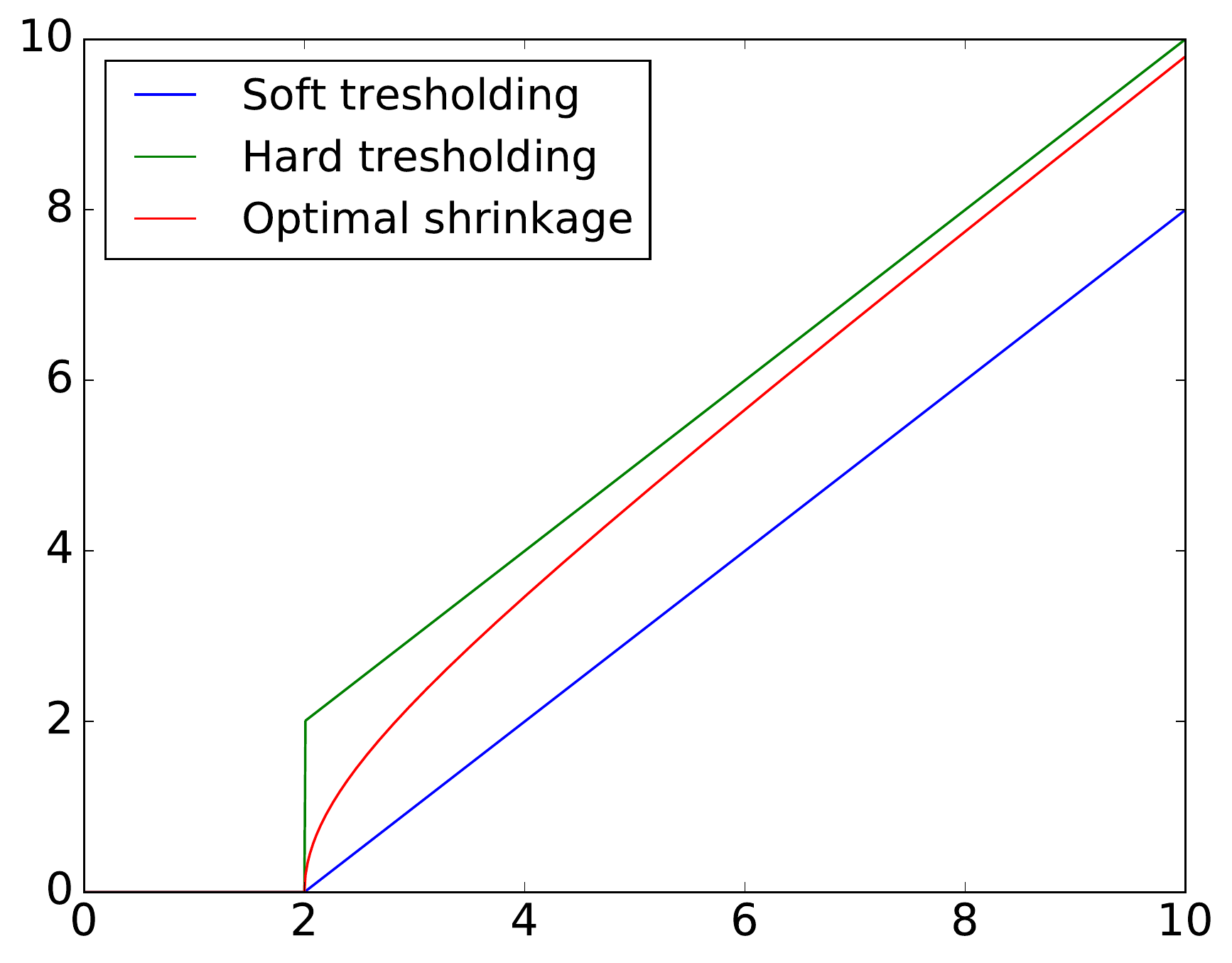}
       \caption{Three common singular value shrinkers. \AMPSVST~ is based on
	      iterative soft thresholding \qref{soft:eq}, whereas \AMPOPT~ uses
	      a variation of the optimal 
	      shrinker given in \qref{opt:eq}. Hard thresholding is used 
       in the iterative hard thresholding family of algorithms, see e.g \cite{tanner2013normalized}. 
       All the plotted shrinkers are tuned for $\beta=1$. 
       }
       \label{shrinkers:fig}
\end{figure}

The main discovery reported here is that by plugging the soft thresholding
singular value shrinker $\eta^{soft}_\lambda$ 
and the optimal singular value shrinker $\eta^{opt}$
into the Matrix AMP iteration
\qref{AMP:eq}, {\em one obtains matrix recovery algorithms which, to the best of
       our knowledge, 
meet or exceed the state-of-the-art }
 in the case of
random linear measurements,
in the following two aspects:

{\bf (i) Convergence rate.} Whenever recovery is possible, 
	      our algorithms typically converge {\em exponentially fast} 
	      in the number of iterations,
	      a phenomenon that has been
documented in AMP iterations for vector recovery
\cite{donoho2009message,donoho2013accurate}. 
We bring substantial evidence that the convergence rate of our algorithms
compares very favorably with state-of-the-art matrix recovery methods, including
first order algorithms based on variations of \qref{GD:eq}.

\begin{figure}[h]
       \centering
       \includegraphics[width=\columnwidth]{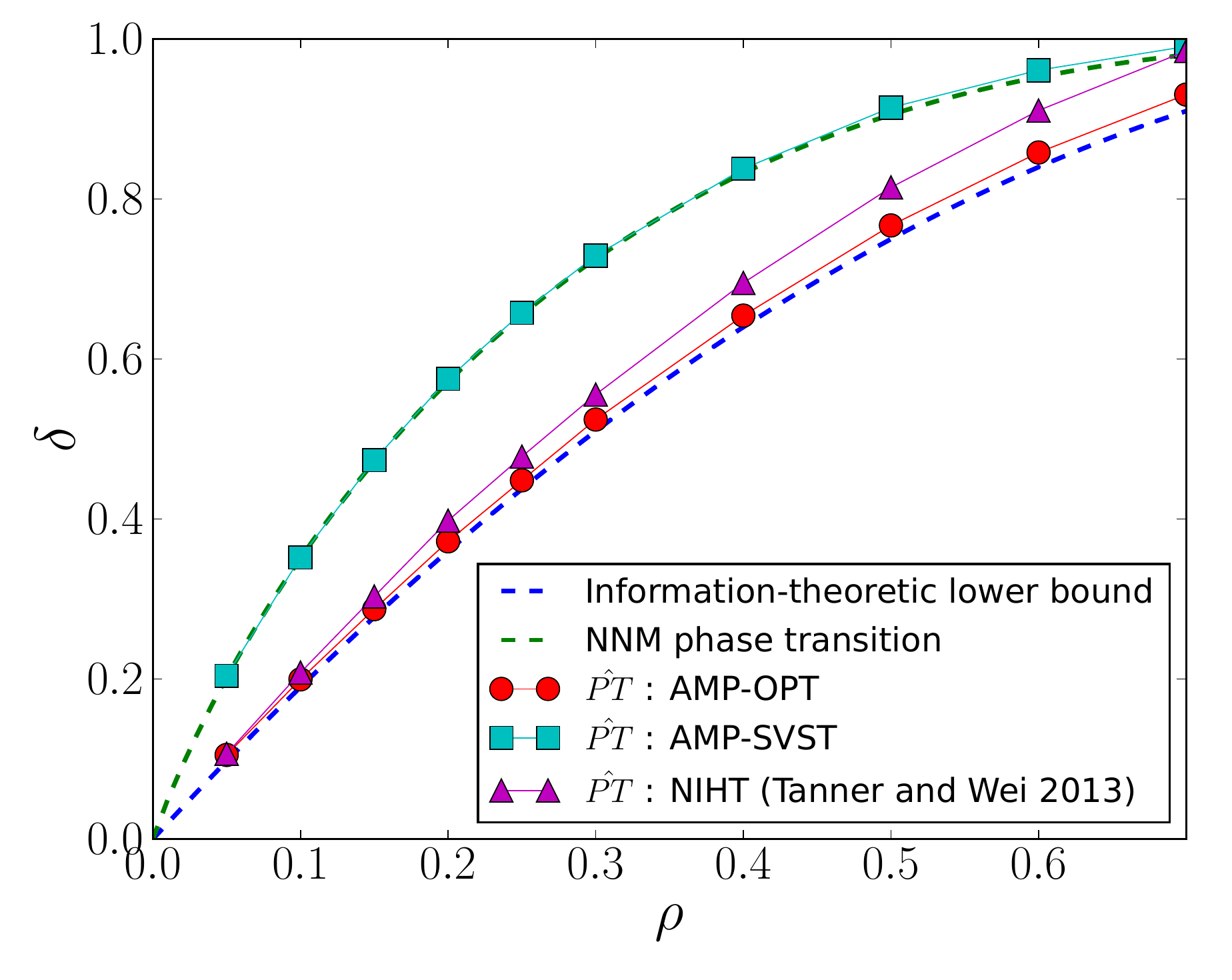}
       \caption{
       Evaluated phase transition of the proposed algorithms \AMPSVST~ and
       \AMPOPT; evaluated phase transition of state-of-the-art 
       algorithm \TW;
       information theoretical lower bound \qref{lower_it:eq}; and
       theoretical
       asymptotic phase transition of NNM
       \qref{NNM_PT:eq}, explicitly 
       provided in the
       SI Appendix, section 5. 
       Gaussian observations, $N=M=100$ ($\beta=1$). See Methods section below
       for details. Similar figures for other values of $\beta$ appear in the
       SI Appendix, section 7.3.
       }
       \label{fig1}
\end{figure}

{\bf (ii) Number of measurements required.}
Figure \ref{fig1}, which summarizes results of 
massive computer simulations described
below, shows that \AMPOPT~ 
 requires a
 near-optimal number of measurements for successful recovery, as we now
 elaborate. 
 To give a concrete example, when $r=rank(X)$ is low, \AMPOPT~ requires
 $\approx r(N+M)$ measurements, which compares favorably with the $\approx 3r(N+M)$ measurements
 required by the popular Nuclear Norm Minimization 
 algorithm, and agrees with the information-theoretic lower bound.

\subsection*{Phase transitions in matrix recovery}
~How many measurements $n$ are required for a given matrix recovery algorithm
to correctly recover a matrix $X$ with $rank(X)=r$? Certainly as $r$ grows (so
that the information content of the object to be recovered grows), $n$ must grow
as well. 
Many variations of \qref{GD:eq} solve the {\em Nuclear Norm
Minimization} (NNM) convex program
\begin{eqnarray*}
      \text{min}_W \norm{W}_*\\
      \text{subject to}\, \Ac(W)=y\,,
\end{eqnarray*}
where $\norm{\cdot}_*$ denotes the nuclear norm of a matrix, namely, the sum of
its singular values. 
In the Gaussian measurements model, several authors have noted a phase
transition phenomenon for NNM, 
whereby for a given value of $r=rank(X)$ there is a fairly
precise number of samples $n(r,M,N)$ such that NNM from $n$ measurements 
typically succeeds if 
$n>n(r,M,N)$ and typically fails otherwise
\cite{recht2010guaranteed,donoho2013phase,oymak2016sharp,amelunxen2014living}.
Following these authors,
we  consider a sequence of 
recovery problems obeying a proportional growth model as $N\to\infty$,
\begin{equation} \label{growth:eq}
	\frac{M}{N} \to \beta \in (0,1], \quad \frac{r}{M} \to \rho \in (0,1],
	\quad \frac{n}{NM} \to \delta \in (0,1]\,,
\end{equation}
where $\beta$, $\rho$ and $\delta$ are known as the aspect ratio, rank fraction
and undersampling ratio, respectively. 
Letting 
$\delta^*(r,M,N)=n(r,M,N)/(MN)$ be the location of the phase transition
 we consider the formal limit 
 \begin{equation} \delta^*(\rho,\beta)
	= \lim_{N\to\infty} \delta^*(N,M,r)\,. 
\end{equation} 

While most results in the literature focus on NNM, it is tempting to study any
matrix recovery algorithm using the same lens, and compare competing algorithms
using their phase transition curves, an approach that has proven very successful
in compressed sensing \cite{donoho2005sparse,donoho2009observed,donoho2011noise,maleki2010optimally}.

\subsection*{Phase transitions and minimax MSE}

It has been empirically \cite{donoho2013phase} and theoretically
\cite{amelunxen2014living, oymak2016sharp} demonstrated that
 NNM (and any iterative singular value shrinkage  algorithm converging to the
 NNM solution) exhibits a phase transition, which we will denote by 
 $\delta^{\nnm}(\rho,\beta)$.
These authors show that $\delta^{\nnm}(\rho,\beta)$ is given
{\em exactly} by the minimax Mean
Square Error (MSE) of matrix denoising by singular value soft thresholding over
matrices of asymptotic rank fraction $\rho$ and aspect ratio $\beta$. Formally,
 \begin{equation} \label{NNM_PT:eq} \small \delta^{\nnm}(\rho,\beta) =
       \lim_{N\to\infty} \min_\lambda \max_{r(X)\leq \rho M} \frac{1}{MN} \E
\norm{\eta^{soft}_\lambda(Y)-X}_F^2\,.  \end{equation} Here, $Y=X+Z$ with $Z$  a
matrix with i.i.d $\Nc(0,1)$ entries, and $\norm{\cdot}_F^2$ is the squared
Frobenius norm, namely the sum of squares of the matrix entries.  The quantity
on the right hand side has been calculated explicitly as a function of $\beta$
and $\rho$, see \cite{donoho2014minimax,donoho2013phase} and the SI Appendix, section 5.

While \qref{NNM_PT:eq} connects the phase transition of NNM, a convex
optimization algorithm, to the minimax MSE of a matrix denoiser, a similar
connection has been observed and verified for many different AMP iterations.
Specifically, under mild conditions on a shrinker $\eta$, and under the
assumption known as AMP {\em State Evolution},
\cite{donoho2013accurate,bayati2011dynamics} found that the phase transition
$\delta^*(\rho)$ of the AMP iteration, corresponding to $\eta$, matches
$\Mc(\rho|\eta)$, the
asymptotic minimax MSE of denoising a signal with rank-fraction $\rho$ using $\eta$.
Formally, 
\begin{eqnarray} \label{deep:eq} \delta^*(\rho) =
\Mc(\rho|\eta)\,.  \end{eqnarray}
\noindent This fundamental connection has been observed for multiple vector recovery
problems \cite{donoho2013accurate}, with the rank fraction $\rho$ replaced by some other measure of
sparsity or information content.

In this paper we observe that the fundamental connection \eqref{deep:eq} holds for the Matrix
AMP iteration \qref{AMP:eq} with two very different choices of shrinker $\eta$.
When $\eta=\eta^{soft}_\lambda$ (with $\lambda$ optimally tuned), one naturally
conjectures that the phase transition of the corresponding Matrix AMP iteration
will exactly match the asymptotic minimax MSE of matrix denoising by
$\eta^{soft}_\lambda$, namely, will have the same phase transition curve as NNM,
given by \qref{NNM_PT:eq}. 
While the literature points to a deep connection between 
 decision theory (specifically, minimax MSE) and compressed
 sensing phase transitions in matrix recovery
 \cite{donoho2013phase,amelunxen2014living,oymak2016sharp},
this observation is, to the best of our knowledge,
the first instance where such connection is made direct and explicit,
in the same sense that  \cite{donoho2009message} has identified a similar connection
in sparse vector recovery problems.
%
%
\subsection*{Matrix denoising and optimal singular value shrinkage}

The identity \eqref{deep:eq} as it holds for AMP-based matrix recovery algorithms suggests that
by designing a singular value shrinker with lower asymptotic 
minimax MSE, 
one might obtain a matrix recovery algorithm with an improved  phase transition.
This can hypothetically be achieved by plugging in the improved shrinker 
into \qref{eta:eq} and using the AMP iteration \qref{AMP:eq}.

Interestingly, in an asymptotic model of low-rank matrices observed in white
noise, 
simple, closed-form formulae are available for the optimal singular value
shrinker \cite{gavish2014optimal,shabalin2013reconstruction}. The optimal singular value shrinker can be shown to be 
\begin{eqnarray} \label{opt:eq}
	\eta^{spiked}(y) = 
	\begin{cases}
		\frac{1}{y}\sqrt{(y^2 -\beta -1)^2 -4 \beta} & y \ge 1+\sqrt{\beta} 
		\\
		0 & y \le 1+\sqrt{\beta}
	\end{cases} \,. 
\end{eqnarray}
Adapting the shrinker $\eta^{spiked}$ to the proportional growth model
\qref{growth:eq} requires technical detail and is deferred to
the SI Appendix, see section 4.2.
We denote the adapted version of \qref{opt:eq}, formally defined in \qref{eta_opt:eq}
below, by $\eta^{opt}$.
The shrinkers $\eta^{opt}$ and $\eta^{soft}_\lambda$ are shown in Figure
\ref{shrinkers:fig} above.

Now, if indeed the adapted version of  \qref{opt:eq} has an appealing worst-case
(hence minimax) MSE, one could hope, according to \qref{deep:eq}, that the corresponding AMP iteration would
have an appealing phase transition.

\subsection*{Unification of lower bounds}

There is reason to believe
that the minimax MSE of the $\eta^{opt}$ is
not only appealing, but in fact near-optimal. 
Indeed, \cite{donoho2014minimax} show that
for {\em any} measurable matrix 
denoiser $\hat{X}:\R^{M\times N} \to \R^{M\times N}$ (not necessarily based on
singular value shrinkage) the
minimax MSE of $\eta$ is lower-bounded by
\begin{equation} \label{lower_MSE:eq}
       \small
		\max_{rank(X)\le r} 
	 \frac{1}{NM}
	 \E  
	 \norm{ X-\hat{X}(X+Z) }_{F}^2  
	 \ge 
	 \frac{r}{M} + \frac{r}{N} - \frac{r^2 + r}{MN}\,.
\end{equation}
Moreover, \cite{donoho2014minimax} show that this lower bound, if achieved, is achieved by a 
singular value shrinker. 
Empirical evidence, included in the SI Appendix (see section 4.3), suggest that the
minimax MSE of $\eta^{opt}$ is close to 
the lower bound \qref{lower_MSE:eq}.  

The connection \qref{deep:eq}
suggests
that 
no Matrix AMP  iteration can achieve a phase transition lower than the limit 
of \qref{lower_MSE:eq}, namely
\[
       \lim_{N\to\infty}  \frac{r}{M} + \frac{r}{N} - \frac{r^2 + r}{MN}
       = \rho(1+\beta-\beta\rho)\,,
\]
and that the Matrix AMP iteration based on $\eta^{opt}$ would have a
 phase transition close to the best phase transition achievable by any matrix
 AMP iteration. 

Surprisingly, simple dimension considerations imply that no matrix recovery (AMP
or other) can have better uniform guarantees:
 there is a simple information-theoretic lower bound on the
phase transition one could possibly hope for in a matrix
recovery algorithm, which we denote by $\delta^{\IT}(\rho,\beta)$.
%
Indeed, as the set of rank-$r$ $M$-by-$N$ matrices is a 
smooth manifold of dimension $r(M+N-r)$ embedded in $\R^{M\times N}$
(see, e.g, \cite{vandereycken2013low}),
faithful recovery of any matrix in the manifold
requires \emph{at least} $n \ge r(N+M-r) $ linear measurements. 
It follows that 
\begin{equation} \label{lower_it:eq}
	\delta^{\IT}(\rho,\beta) = 
	\lim_{N\to\infty} \frac{r(N+M-r)}{NM} = \rho(1 + \beta - \beta\rho) \,.
\end{equation}
This 
lower bound agrees, to order $1/N$, with \qref{lower_MSE:eq}, which stems from
an altogether different consideration.
In other words, if a Matrix AMP iteration based on $\eta^{opt}$ 
is shown to have a phase transition close to the curve
$\delta(\rho)=\rho(1+\beta-\beta\rho)$, as predicted by 
\qref{deep:eq},
then it is in fact near-optimal among {\em any} matrix
recovery algorithm whatsoever, in terms of the number of measurements required
for successful recovery.



\subsection*{Detailed description of the proposed algorithms}

Complete specification of the proposed algorithms requires full specification of 
$\nabla\cdot \eta_t$ from \qref{AMP:eq}, as well as $\hat{\sigma}_t$ and $\eta$
from \qref{eta:eq}. 
To estimate the noise level $\hat{\sigma}_t$ we can use
\cite{donoho2009message,donoho2013accurate} 
\begin{equation}
       \hat{\sigma_t} = \frac{1}{\Phi^{-1}(0.75)}median(\abs{z_t}) \,,
\end{equation}
where $z_t$ is defined in \qref{AMP:eq}.
The divergence term in \qref{AMP:eq} for a general
singular value shrinker is given by \cite{candes2013unbiased,donoho2014minimax}
{\small
\begin{equation}
\label{divergence_general:eq}
\begin{split}
	\nabla \cdot \eta(X) 
	&= \sum_{i=1}^M \frac{d\eta}{d\sigma}(\sigma_i)
	+ 2 \sum_{i<j} \frac{\eta(\sigma_i)\sigma_i - \eta(\sigma_j)\sigma_j}{\sigma_i^2 - \sigma_j^2} \\
	&+ (N-M) \sum_{i=1}^M \frac{\eta(\sigma_i)}{\sigma_i}
\end{split}
\end{equation}
}
where $\sigma_1> \ldots > \sigma_M>0$  are the singular values of the matrix
argument $X$, whose spectrum is assumed to be non-degenerate. 
(Indeed, degenerate spectrum occurs with probability zero in any matrix model of
interest).
The derivative $d\eta/d \sigma$ is shown next for each choice of shrinker.

\paragraph{Algorithm 1: \AMPSVST.}
We use $\eta^{soft}_\lambda$ \qref{soft:eq} with a tuning parameter
$\lambda = \sqrt{N}\lambda^*(\rho,\beta)$ achieving the 
asymptotic minimax MSE in the right hand
side of \qref{NNM_PT:eq}. Details of calculating $\lambda^*(\rho,\beta)$ are
deferred to the SI Appendix (see section 5).
The (weak) derivative of $\eta^{soft}_{\lambda}$ is simply
     $  d\eta^{soft}_{\lambda}/d \sigma =
     \mathbf{1}_{\{\sigma>\lambda\}}$\,.

\paragraph{Algorithm 2: \AMPOPT.}
We use a properly calibrated version 
of $\eta^{spiked}$ from \qref{opt:eq}, defined by
{\small
\begin{equation} \label{eta_opt:eq}
       \eta^{opt}(\sigma) = \alpha\sqrt{N} \cdot \eta^{spiked} 
       \left( \frac{\sigma}{ \alpha\sqrt{N}}\right)\,,
\end{equation}
}
where 
$	
\alpha = 
(\sqrt{1-\beta\rho} + \sqrt{\beta-\beta\rho}) / (1+\sqrt{\beta})
	$.
The full derivation of \qref{eta_opt:eq} and $\alpha$ is deferred to
the  SI Appendix, section 4.2.
Now, the derivative of $\eta^{spiked}$ from \qref{opt:eq} is given by
{\scriptsize
\begin{eqnarray*}
       \frac{d\eta^{spiked}}{d\sigma}(\sigma) = 
	    \Bigg(-\frac{1}{\sigma^2}\sqrt{ (\sigma^2- \beta -1)^2 - 4 \beta }
	    + \frac{2\sigma(\sigma^2- \beta -1)}{\sqrt{ (\sigma^2- \beta -1)^2
     - 4 \beta }} \Bigg) \nonumber
\end{eqnarray*}
}
when $\sigma>1+\sqrt{\beta}$,
so that the derivative of $\eta^{opt}$ is given by
$(d\eta^{spiked}/d\sigma) \left(\sigma/(\alpha\sqrt{N})\right)$.

%
       
\subsection*{Main hypotheses} 

This brief announcement tests two hypotheses regarding the merits of our
proposed algorithms, by conducting
substantial computer experiments generating large numbers of random problem
instances. 

\paragraph{Phase transitions.}

Based on \qref{deep:eq} we hypothesize that the phase transition of the
Matrix AMP algorithm \AMPSVST~ matches \qref{NNM_PT:eq}  exactly, and that the 
phase transition of \AMPOPT~ is close to the lower bound  \qref{lower_it:eq}.

\paragraph{Convergence rates.}

One of the appealing properties of AMP algorithms for sparse vector recovery is 
their exponential rate of convergence \cite{donoho2009message, donoho2013accurate}
-- see SI Appendix section 6.3 for more details. We hypothesize
that when $NM$ is large both \AMPSVST~ and \AMPOPT~ similarly 
demonstrate an exponential rate of convergence
whenever recovery is possible.

\section*{Methods}
{
       \small

We use statistical methods to check for agreement between the
hypothesis and the predicted phase transition. We further compare both phase
transitions and convergence rates of our proposed algorithms 
with two different state-of-the-art algorithms
for matrix recovery.

\paragraph{Comparison with state-of-the-art.}       

The Matrix AMP framework we propose, and in particular the proposed algorithms
\AMPSVST~ and \AMPOPT, are essentially iterative
singular value thresholding methods. We therefore compare them with 
well-known
algorithms based on variations of iterative  singular value shrinkage, 
\eqref{GD:eq}. (SI Appendix section 7.5  also compares \AMPOPT~ with an involved, 
state-of-the-art AMP algorithm, developed from first principles, which has a
completely different form.) In order to compare convergence rates of the
algorithms close to their respective phase transitions, 
we match \AMPSVST~ (resp. \AMPOPT) with an algorithm whose phase transition is
expected to be similar.

\begin{enumerate}
       \item {\em Accelerated proximal gradient singular value thresholding} 
	      (\TY) \cite{toh2010accelerated}: An accelerated proximal gradient descent algorithm
	      for NNM, based on soft thresholding of singular values. As an NNM
	      solver, \TY~ is expected to have the same phase transition as
	      \AMPSVST. Being a
	      simple, general-purpose solver, it serves as a reasonable baseline
	      for our evaluation\footnote{Some related gradient
		   descent algorithms have been shown to
		   converge exponentially fast, 
		   e.g. \cite{agarwal2012fast}. However they
		   sometimes fail to converge to the NNM solution
		   \cite{oymak2015sharp} and
		   are therefore not suitable candidates for comparison to
		   \AMPSVST.
	      }.
	      
       \item {\em Normalized iterated hard thresholding} (\TW) \cite{tanner2013normalized}: An 
	      alternating projection algorithm, with an adaptive stepsize,
	      based on hard thresholding of singular values. In
	      \cite{tanner2013normalized} the authors bring some evidence that
	      this algorithm exhibits a phase transition which is not far from
	      the information-theoretical lower bound. To the best of our
	      knowledge, its convergence rate is also state-of-the-art.
	      %
	      %
\end{enumerate}
Both algorithms are described in detail in the
SI Appendix, section 3.
We studied both the computational complexity and the phase transition of all
four algorithms, as follows.

\paragraph{Evaluating convergence rates.}

All four algorithms under study are iterative, with roughly the same
per-iteration complexity. 
To compare their overall complexity, we measured the rate of convergence of each
algorithm. 
We recorded the relative error 
$\Delta_{t} := \norm{X_t-X}_F/\norm{X}_F$ of each algorithm under study 
over the number of iterations $t$,
with $X$ being the true  (unknown) matrix to be recovered.
Due to space considerations, a detailed account of our methodology
is deferred to the SI Appendix, see section 2.

\paragraph{Evaluating phase transitions.}

Our analysis follows the methods of 
\cite{donoho2009observed, monajemi2013deterministic, donoho2013phase}. 
For each of the proposed algorithms we assigned a suspected phase
transition $\delta^*$. The suspected phase transition for 
\AMPSVST~ was $\delta^* = \delta^{\nnm}$ from \qref{NNM_PT:eq},
and for \AMPOPT~ and \TW~ we used $\delta^* = \delta^{\IT}$ from \qref{lower_it:eq}.
For several problem dimensions $N$ and parameter combinations $M=\beta N,r=\rho M$ we
performed many random recovery experiments, using undersampling ratios $\delta$ above and below 
$\delta^*$. 
We used matrices with fully degenerate spectrum, 
a configuration known to be least favorable for \AMPSVST~ but possibly not for
\AMPOPT~ (see important
discussion in the SI Appendix, section 6.2).
At each point in parameter space, we chronicled the empirical probability of success $\pi(r|n,N,M)$. 
We then fitted a logit 
curve $\log(\hat{\pi}/(1-\hat{\pi}))=a+b(\delta-\delta^*)$. Setting $\pi=1/2$, we obtain an estimate 
$\hat{\delta} = \delta^* - a/b$ of the real phase transition at $\rho,\beta$.  
Due to space consideration, we defer further details of our methodology to the SI Appendix (see section 2). 
\ifx\MASS\undefined\else{

\paragraph{Evaluating convergence rates.}
All four algorithms under study are iterative, with roughly the same
per-iteration complexity. 
To compare their overall complexity, we measured the rate of convergence of each
algorithm. 
We used Gaussian measurements with parameter values 
$N=M=50$ ($\beta=1$) and $N=100$, $M=50$ ($\beta=0.5$), with 
$\rho\in \left\{ 0.04, 0.1, 0.2, 0.4 \right\}$. 
In one set of experiments the number of measurements was
$n=\lceil(0.03+\delta^*)\cdot MN \rceil$
(undersampling ratio $\delta$ close to the suspected phase transition);
in another set we used $n=\lceil (0.1+\delta^*)\cdot MN \rceil$ (far
from the suspected phase transition). 
Here, $\delta^*=\delta^{\IT}$ from \qref{lower_it:eq} for \AMPOPT~ and \TW,
and $\delta^*=\delta^{\nnm}$ from \qref{NNM_PT:eq} for \AMPSVST~ and \TY.
We performed $B=50$ monte carlo simulations for each algorithm and combination of parameter values.
In each monte carlo simulation $b$ ($1\leq b\leq B$) a problem instance was
generated (see below) and the algorithm was initialized.  The algorithm was
then allowed to run
for $T\ge 500$ iterations without interruption, 
and at each iteration the relative error
\[
       \Delta_{t,b} := \frac{\norm{X_t-X}_F}{\norm{X}_F}
       \]
was recorded, with $X$ being the true  (unknown) matrix to be recovered.
The $B/2$ worst trials (for which $\Delta_{T,b}$ was maximal) were then discarded
and the remaining values were averaged to produce the monte carlo averages
$\overline{\Delta}_t:= \frac{1}{B/2}\sum_{b: \Delta_{T,b}\le \text{median}} \Delta_{t,b}$. 
This allowed us to study the behavior of the algorithm during a
typical \emph{successful}
run, see discussion in the SI Appendix, section 2. 
}
\fi

 \ifx\MASS\undefined\else{

\paragraph{Evaluating phase transitions.}
Our analysis follows the methods of 
\cite{donoho2009observed, monajemi2013deterministic, donoho2013phase}. 
For each of the proposed algorithms we assigned a suspected phase
transition. The suspected phase transition for 
\AMPSVST~ was $\delta^{\nnm}$ from \qref{NNM_PT:eq}, see also SI Appendix, section 5.
The suspected phase 
transition of  \AMPOPT~ was $\delta^{\IT}$ from \qref{lower_it:eq}.
For some parameter values, we also studied the algorithm \TW~ in a similar manner with suspected phase
transition  $\delta^{\IT}$ as above (this was done for much fewer parameter combinations). 
Since \TY~ converges to the NNM solution, 
its phase transition is known to be equal to  $\delta^{\nnm}$, (see e.g.
\cite {donoho2013phase})
       and was not
studied. 
%
Define the probability of correct recovery of a target matrix $X$ by 
a given iteration $X_t$ as follows
\begin{eqnarray}
       \pi(r|n,N,M) = Pr\left\{ X = \lim_{t\to\infty} X_t \right\}\,.
\end{eqnarray}
To estimate 
$\pi(r|n,N,M)$ we used $B\geq 50$ identical monte carlo simulations. In each
simulation we generated a problem instance (see below), initialized the
algorithm and ran it for $T$ iterations. In each iteration we measured the
relative error 
$ \norm{X_t-X}_F / \norm{X}_F$, using oracle access to the true value of $X$.
If the relative error was less than $\epsilon = 10^{-3}$
before iteration $T$ was reached, the monte carlo simulation was considered a
successful recovery. Otherwise, it was considered a failed recovery. We recorded
$\hat{\pi}(r|n,M,N)=\#\text{successes}/B$ for each of the algorithms under study, 
where we sampled values of $\delta$ in a window around $\delta^*$ (the suspected
phase transition) with equal spacings $0.01$ (typically
in the range $[-0.05,0.05]+\delta^*$, in some experiments the interval extended further beyond $\delta^*$). 
We then proceeded to estimate the location of the phase transition for each
algorithm by fitting a logit function
$\log(\hat{\pi}/(1-\hat{\pi}))=a+b(\delta-\delta^*)$ where $\delta^*$ is the
location of the suspected phase transition at $M$,$N$ and $n$.
For each algorithm, we plot $\hat{\pi}$ over $\delta$ and estimate the location
of the phase transition by $\delta^*-a/b$, an extrapolation for the value of
$\delta$ such that 
$\hat{\pi}(\delta)=1/2$.
We performed an extensive set of experiments for comparison of phase
transitions: (i) low-rank experiments with $N=100$, $T=4000$ and  $\rho$ ranged from $0.05$ to
$0.3$ in increments of $0.05$; this was performed for $\beta\in\left\{
       0.5,1
\right\}$.
(ii) higher-rank experiments with $N \in \left\{ 50,100,150 \right\}$, $T=4000$ where $\rho$ ranged
from $0.1$ to $0.6$ in increments of $0.1$, and $\beta\in\left\{
       0.2,0.4,0.6,0.8,1.0
\right\}$. 
In each experiment we plotted the estimated phase transition $\delta(\rho)$ over
$\rho$.
}
\fi

\ifx\MASS\undefined\else {

\paragraph{Generation of problem instances.}
Each problem instance $(\Ac,y)$ for parameters $(M,N,r,n)$ 
was generated as follows. We first generated 
a pseudo-random matrix $X=\mu UV'$ where $U$ is $M$-by-$r$ and $V$ is $N$-by-$r$, both
uniformly distributed on their respective Stiefel manifolds. $\mu=100$ was used
in most cases. We then generated a measurement matrix $\Ac$ distributed  i.i.d
$\Nc(0,1/n)$ and let $y=\Ac(X)$. As discussed in the SI Appendix, section 6.2,
performance of the Matrix AMP algorithms is invariant to the choice of $\mu$. As
discussed there, while  
signals with  degenerate spectrum $X=\mu UV'$ are least favorable for \AMPSVST,
the least favorable spectrum configuration for \AMPOPT~ is unknown. 
Extension of our study to additional spectral configurations is however beyond
the scope of this brief announcement.
}
\fi

 \ifx\MASS\undefined\else{
\paragraph{Universality.}
To test whether our results depend on the particular distribution used to
generate the entries of the measurement matrix $\Ac$ we performed local
evaluation of the phase transition of \AMPSVST~ and \AMPOPT, for parameters
$\rho\in \left\{ 0.1,03 \right\}$ and $\beta\in \left\{ 0.5,1.0 \right\}$ (with $N=100$), using several distributions: Gaussian (light tail), 
Rademacher ($\pm1$ with equal probability -- no tail) 
and Student-t with $\nu=6$ degrees of freedom (heavy tail).
All
distributions are symmetric about $0$, and were normalized to have variance $1/n$.
}
\fi

\paragraph{Simulation software platform.}
Empirical evaluation of matrix recovery
phase transitions requires computer simulation on a massive scale, which poses a
software development challenge. 
To efficiently conduct the simulation reported in this paper, 
we developed a dedicated 
software platform inspired by \cite{donoho2013phase,monajemi2013deterministic}.
Our framework is written in
{\tt Python} and is made available on the Data and Code Supplement \cite{SDR}.
We used the {\tt Spark} parallelization framework
\cite{zaharia2010spark} to orchestrate 
the parallel computations required, and executed the code on large Amazon Web
Services machines using roughly $15,000$ CPU hours. 

}

\renewcommand{\thefootnote}{$\dagger$} 
\section*{Results
\footnote{{\em Reproducibility advisory.} 
All the figures and tables in this paper, including those in the
SI Appendix, are fully reproducible. 
Raw simulation results were stored in a {\tt mysql} database from which all
figures were generated.
All our scripts, as
well as a snapshot of the full result database, is available
in the Data and Code Supplement 
\cite{SDR}.}}
\begin{figure}[h]
       \centering
       \includegraphics[width=\columnwidth]{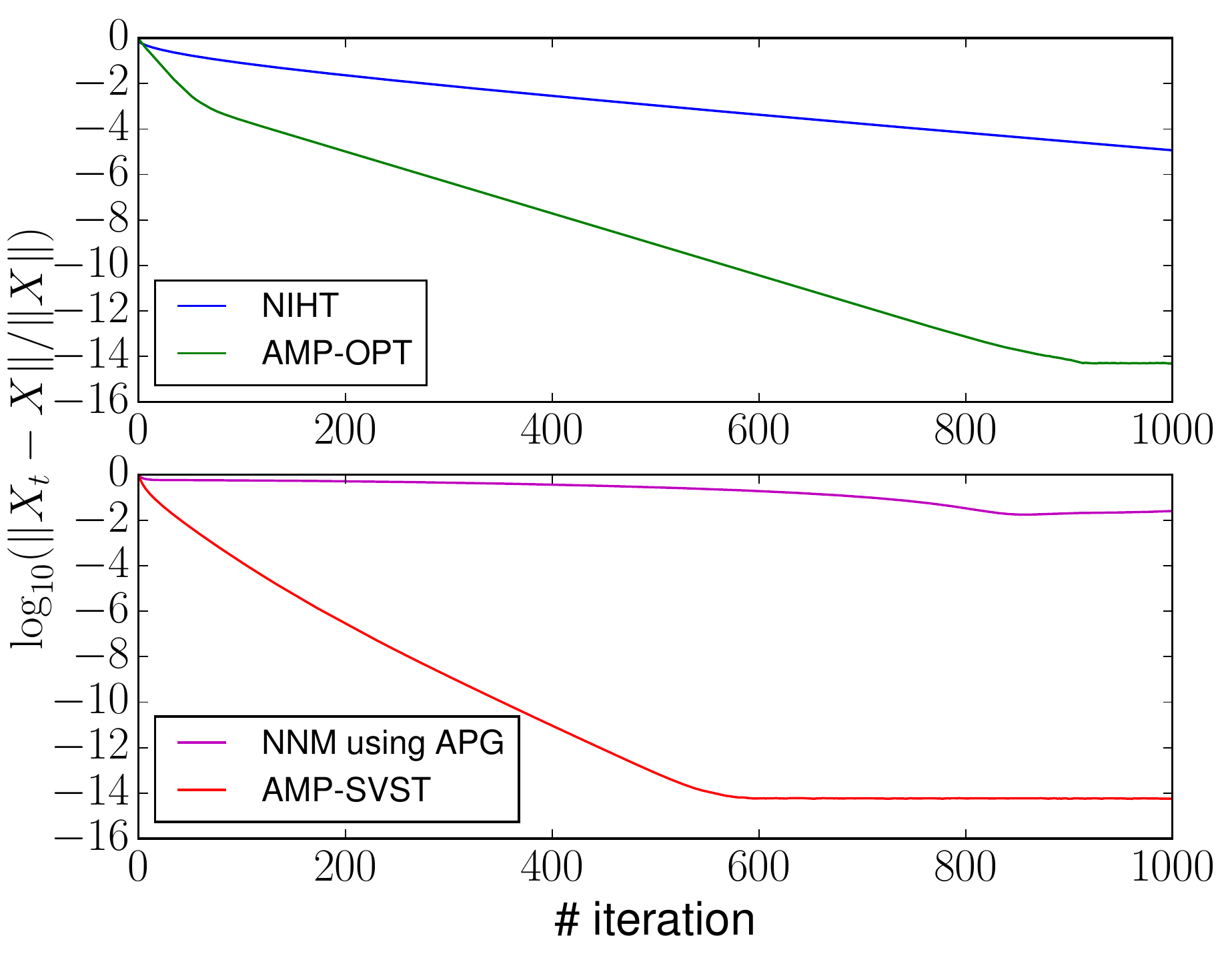}
       \caption{Comparison of convergence rates: relative error (log scale) 
	      over iteration
       number. Both
matrix AMP algorithms exhibit exponential convergence in iteration number until
machine precision is reached (\TW~ appears to converge exponentially as well,
yet significantly slower). See methods section for further details. $N=M=50$
(so $\beta=1$) and
$\rho=0.2$. 
Top: \TW~ and \AMPOPT~ at $\delta=\delta^{it}(\rho,\beta)+0.1$.
Bottom: \TY~ and \AMPSVST~ at  $\delta=\delta^{nnm}(\rho,\beta)+0.1$.
}
       \label{fig2}
\end{figure}

\begin{figure*}[h]
       \centering
       \includegraphics[width=0.75\columnwidth]{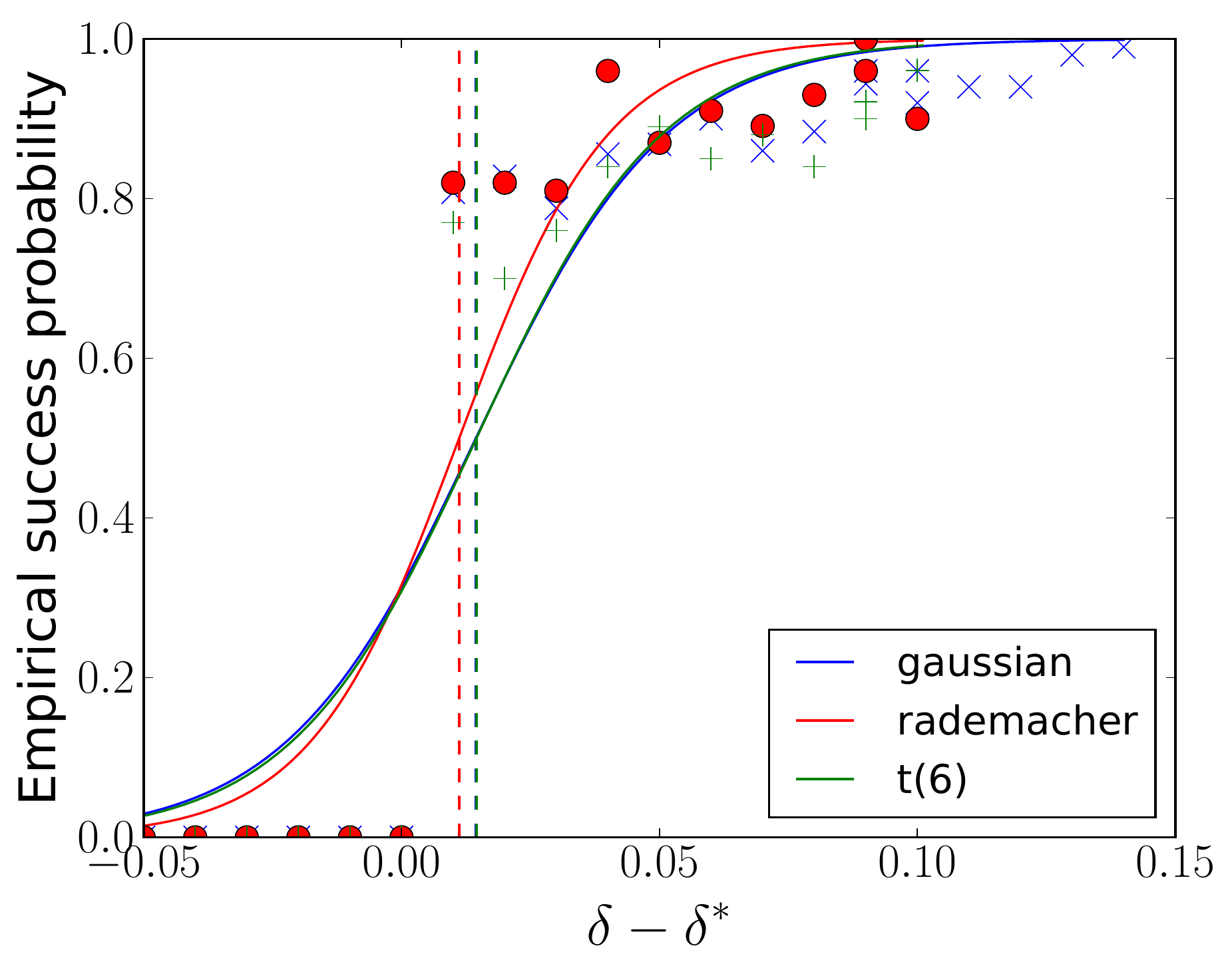}
       \includegraphics[width=0.75\columnwidth]{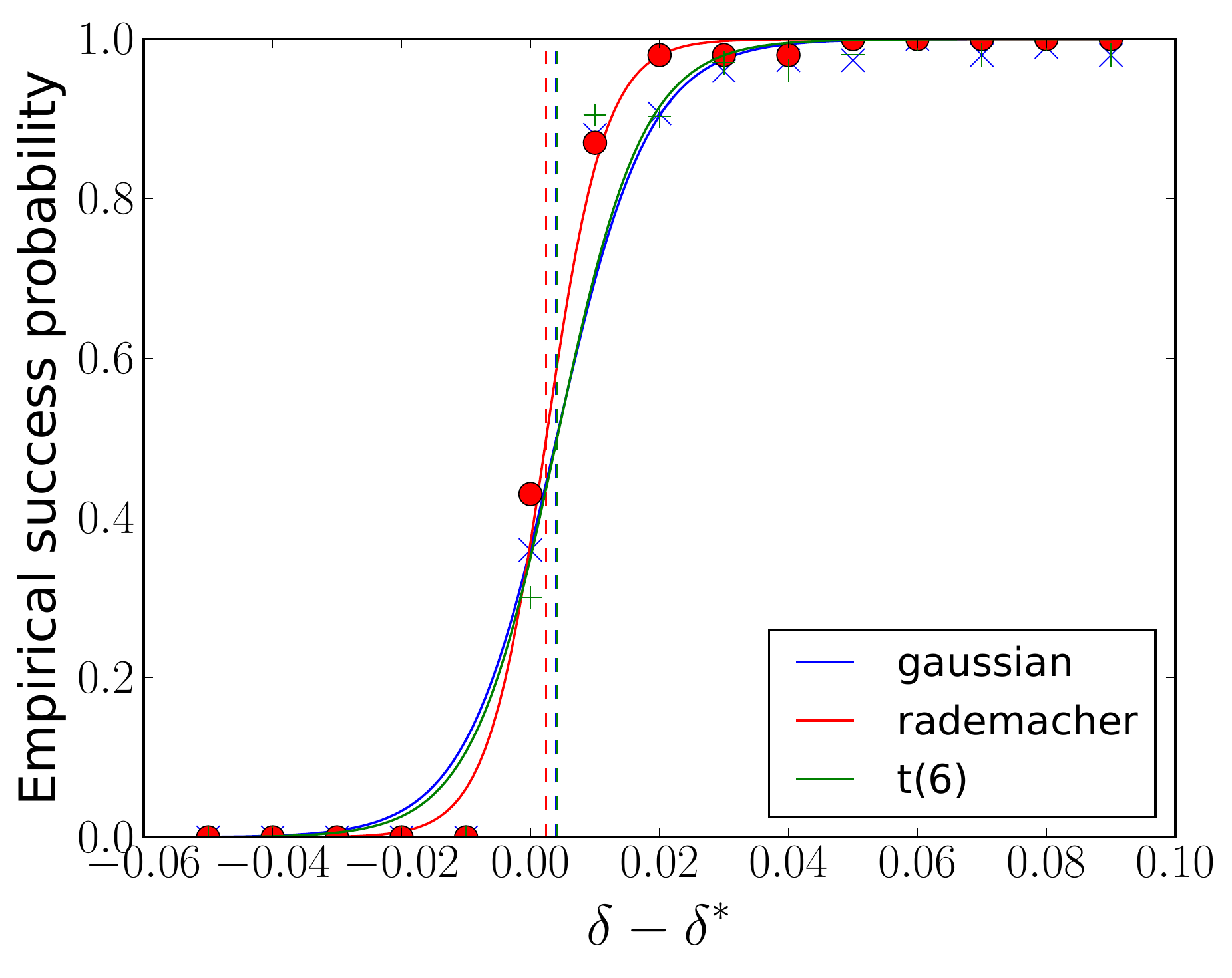}
       \caption{Evidence of universality. Empirical success probability (with
       logit fit and estimated phase transition) for various probability
       distributions of the entries of the measurement matrix $\Ac$. 
       Left:  \AMPOPT. Right: \AMPSVST. $M=N=100$, $\rho=0.3$.   }
       \label{fig:universality}
\end{figure*}

\paragraph{Convergence rates.}
Figure \ref{fig2} shows typical convergence rates of the algorithms under
study above their respective phase transitions for $\rho=0.2$ and $\beta=1$ 
contrasting \AMPSVST~ with \TY~ (at $\delta=\delta^{nnm}(\rho,\beta)+0.1$, close to
their common phase transition) and \AMPOPT~ with \TW~ (at
$\delta=\delta^{it}(\rho,\beta)+0.1$, close to the information-theoretical lower bound).
When successful, matrix AMP algorithms exhibit roughly exponential convergence rates;
for example, after $T=1000$ iterations, the relative error for
\AMPOPT~  is $10^{-14}$, compared with 
$10^{-5}$ for \TW.  Results for
additional values of $\rho$ and $\beta$ are deferred to the SI Appendix (see section 7.3). Our results support the hypothesis that Matrix AMP algorithms
converge exponentially fast when recovery is possible.

\paragraph{Phase transitions.}
Figure \ref{fig1} shows the estimated phase transition of the algorithms
\AMPSVST, \AMPOPT, \TW, as well as the asymptotic phase transition of \TY~ 
for $\beta=1$ and $N=100$.
Similar results for $\beta=0.5$, $N=100$ appear in the
SI Appendix, section 7.2 (for all three algorithms above), and for $N\in \left\{ 50,100,150 \right\}$
and $\beta\in \left\{ 0.2,0.4,0.6,0.8,1.0 \right\}$ (for \AMPOPT~ and \AMPSVST).
We found that \AMPSVST~ exhibits a sharp phase transition, matching the phase
transition of NNM and \TY. Our results conclusively affirm 
the relation \qref{deep:eq} for \AMPSVST.
We further 
found that \AMPOPT~ exhibits a phase transition that is less pronounced that
previously studied AMP algorithms, see Discussion below. 
These results support the hypothesis that the phase transition of \AMPOPT~
lies very close to the information-theoretic lower bound \qref{lower_it:eq}.

 \ifx\MASS\undefined\else{
\paragraph{Universality.}
Figure \ref{fig:universality}  shows 
the empirical success probability of \AMPOPT~ and \AMPSVST~ for a 
single choice of parameters, $\beta=1.0$ and $\rho=0.3$ (for $N=100$). 
The results of similar experiments for $\beta\in \left\{ 0.5,1.0 \right\}$ and $\rho\in \left\{ 0.1,0.3 \right\}$ 
($N=100$) appear in the SI Appendix, section 7.4. We consistently observed that the phase transitions for all
different
matrix ensembles under study agree to within measurement error.
}
\fi


\section*{Discussion}

\paragraph{Our contributions.}
\begin{enumerate}
       \item We present overwhelming evidence that 
	      the AMP framework, previously studied for sparse vector recovery and
	      related problems, naturally extends to matrix recovery,
	      with the 
	      notion of matrix rank replacing that of vector sparsity.
	      We validate the correspondence between the phase  transition of
	      \AMPSVST~
	      and the minimax MSE of the underlying shrinker
	      $\eta^{soft}_\lambda$.
	      Our results regarding both \AMPSVST~ and \AMPOPT~ 
	      suggests Matrix AMP as an appealing framework for designing
	      matrix recovery algorithms.
	      
       \item  We present two algorithms, \AMPSVST~ and \AMPOPT, which
	      converge exponentially fast
	      when they succeed.  	      
	      To the best of our knowledge, these algorithms converge at a rate 
	      which meets or exceeds the state-of-the-art in matrix recovery
	      from random linear measurements, sometimes by a significant margin.
	      As discussed below, 
	      in contrast with previously suggested iterative algorithms for
	      matrix recovery, 
	      both \AMPSVST~ and \AMPOPT~
	      offer clear ``diagnostic'' for whether recovery was successful.

       \item We show that while algorithm \AMPOPT~ improves on the
	      state of the art in terms 
	      convergence rate, it also offers a  near-optimal 
	      phase transition
	      for
	      $N,M\gtrsim 100$ (see important comments below).

\end{enumerate}

\paragraph{Universality.} 
To test whether our results depend on the particular
distribution used to generate the entries of the measurement matrix $\Ac$ we
performed local evaluation of the phase transition of \AMPSVST~ and \AMPOPT~
using several distributions: Gaussian (light tail), Rademacher ($\pm1$ with
equal probability -- no tail) and Student-t with $\nu=6$ degrees of freedom
(heavy tail).  All distributions are symmetric about $0$, and were normalized to
have variance $1/n$.  
For details of this experiment, see SI Appendix, section 7.4.
We observe (see Figure~\ref{fig:universality} and
additional results in the SI Appendix, section 7.4) that the phase
transitions of \AMPSVST~ and \AMPOPT~ are the same, to high precision, under
these three different distributions of the measurements matrix.  Other authors
have previously observed universality in other signal recovery settings:
\cite{donoho2009observed} observed universality in sparse vector recovery using
$l_1$ norm minimization; \cite{donoho2013phase} observed in passing
universality for matrix recovery with NNM; \cite{donoho2013accurate} observed
universality for AMP in various vector recovery-related problems.

\paragraph{State evolution.}
We observe that Matrix AMP with our two denoisers behaves as predicted by the
AMP formalism of \emph{State Evolution} (SE)
\cite{donoho2009message,donoho2013accurate}.
Background and empirical evidence
are presented in the SI Appendix, see section 6.

\paragraph{Other approaches to recovery by Approximate Message Passing.}

The term Approximate Message Passing (AMP) is somewhat overloaded in the
literature. 
Previous works in Approximate Message Passing
algorithms for 
low-rank matrix recovery propose a
generative model for low-rank matrices and derive involved AMP algorithms from
first principles \cite{SSZ16,PS16,PSC14a,PSC14b,KMZ13,MATA13,RF12}.
In contrast, this paper follows the framework of \cite{donoho2009message},

which delineates
simple, highly structured iterative thresholding algorithms, whose design basically boils down to
the design of the denoising nonlinearity. 
In SI Appendix Section 7.5 we demonstrate that \AMPOPT~ compares favorably, for the case of random linear measurements, with 
P-BiG-AMP, a notable AMP algorithm for matrix recovery, derived from first
principles.

\paragraph{Stopping conditions and sign of convergence.}

State Evolution suggests that the estimated noise level $\hat{\sigma}_t$ in
every iteration is in fact a good proxy for the current MSE: $\hat{\sigma}_t^2
\approx \frac{1}{\delta NM} \norm{X_t-X}^2$. In contrast with previously
suggested iterative methods for matrix recovery, Matrix AMP algorithms 
therefore have the
important 
advantage of offering continuous \emph{diagnostic measurements} for the quality of the reconstruction $X_t$, and can choose to stop or continue accordingly.
Background and empirical evidence
are presented in the SI Appendix, section 6.4.

\paragraph{Finite-$N$ width of the \AMPOPT~ phase transition.}
Previously studied AMP algorithms for vector recovery problems demonstrated
that, in finite problems, the transition from failed to successful recovery occurs
in a transition region whose width shrinks with problem size. 
We observed that while the transition region of \AMPOPT~ does shrink with $N$, 
it is wider and shrinks more slowly than the transition region of \AMPSVST~ 
and previously
studied AMP algorithms for vector recovery, 
see for instance Figure \ref{fig:universality}.
Furthermore, evidence included in the SI Appendix  suggests that 
below its phase transition, \AMPOPT~  fails always, while
%
%
above its phase transition it succeeds with probability that approaches $1$
rather slowly as $N$ grows, see SI Appendix Section 7.1.
%
%
We believe this happens because 
the shrinker $\eta^{opt}$ is not Lipschitz continuous (see
Figure \ref{shrinkers:fig})  
\cite{bayati2011dynamics,rush2015capacity,rush2016finite}.

\paragraph{Optimality.} 
We conjecture 
that the information-theoretical lower bound 
$n/(NM) \ge \frac{r}{M}(1+M/N-r/M)$ and 
the denoising lower bound
\qref{lower_MSE:eq} are both asymptotically tight. 
Also, one naturally wonders whether the denoiser $\eta^{opt}$ in \qref{eta_opt:eq} is in
fact optimal in the (asymptotic) minimax sense. 
Numerical evidence suggests that
this is not the case, at least when $\rho$ is sufficiently large - see
SI Appendix, section 4.3. Nonetheless, the worst-case
asymptotic MSE of $\eta^{opt}$ seems to be very close to the minimax MSE.

\paragraph{Implementation with fast SVD.}
 It is natural to use fast randomized SVD methods such as
 \cite{liberty2007randomized} to accelerate the SVD step in the proposed
 algorithms. Note however that these methods  introduce
non-trivial complications and noise sensitivities which are likely to affect the
phase transition.
%

\section*{Conclusion}

This paper presents a near-optimal matrix recovery algorithm based on
the Approximate Message Passing framework (AMP).
The same ideas underlying AMP in previously studied vector recovery problems
have been
shown to hold intact in the matrix recovery problem. In particular, our results
show that the AMP
framework bridges  the seemingly unrelated problems of
matrix denoising on the one hand and matrix recovery from partial measurements
on the other hand.
Indeed, design of a
near-optimal matrix denoiser $\eta^{opt}$ 
has lead to \AMPOPT, a near-optimal matrix recovery
algorithm.
For example, when $r=rank(X)$ is low, \AMPOPT~ requires $n\approx r(N+M)$ measurements for recovery, 
whereas NNM
requires $n\approx 3r(N+M)$. 
Our results 
point to a new approach for designing matrix recovery algorithms, via an
interesting connection with statistical estimation and decision theory.

\section*{Acknowledgement}
     We thank the anonymous reviewers for many helpful comments. 
     This work has been partially supported by Israeli Science Foundation
       grant no. 1523/16 and German-Israeli Foundation for Scientific Research
       and Development no. 
       I-1100-407.1-2015.
       ER was partially supported by the HUJI Leibniz center.

\bibliographystyle{abbrv}
\bibliography{ref}

\end{document}